\documentclass[rnote]{aa}            % referee version: for submission

\usepackage{graphicx}             %for PS/EPS graphics inclusion, new
\usepackage{txfonts}
\usepackage{natbib}
\usepackage{amssymb,amsmath}
\usepackage{multirow}
\bibpunct{(}{)}{;}{a}{}{,}
\usepackage[pagebackref=true]{hyperref}
\hypersetup{pdftitle = HC3N in galaxies, pdfauthor = Xue-Jian Jiang, pdfsubject= The subject, pdfkeywords = keyword1 keyword2 keyword3} 

\begin{document}
\newcommand{\Msun}{\ensuremath{M_\mathrm{\odot}}}
\newcommand{\Lsun}{\ensuremath{L_\mathrm{\odot}}}
\newcommand{\LIR}{\ensuremath{L_\mathrm{IR}}}
\newcommand{\hcop}{HCO\ensuremath{^+}}
\newcommand{\hcccn}{HC$_3$N}
\newcommand{\hhco}{H$_2$CO}
\newcommand{\cch}{C$_2$H}
\newcommand{\kms}{km s$^{-1}$}

\title{HC$_3$N Observations of Nearby Galaxies\,$^*$
\thanks{Based on observations with the 100-m telescope of the MPIfR
(Max-Planck-Institut f\"ur Radioastronomie) at Effelsberg, and the Submillimeter 
Telescope (SMT).
The SMT is operated by the Arizona Radio Observatory (ARO),
Steward Observatory, University of Arizona.}
}
%   \subtitle{I. Place Your Subtitle Here}

%   \volnopage{Vol.0 (201x) No.0, 000--000}      %%preserved for Editor. DOn't remove!
%   \setcounter{page}{1}          %%starting page, preserved for Editor. DOn't remove!

   \author{Xue-Jian Jiang
      \inst{1,5}
   \and Jun-Zhi Wang
      \inst{2}
   \and Yu Gao
      \inst{1,5}
   \and Qiu-Sheng Gu
      \inst{3,4,5}
   }
\institute{Purple Mountain Observatory \& Key Laboratory for Radio Astronomy,
  Chinese Academy of Sciences, 2 West Beijing Road, Nanjing 210008, China;
  Email: {\it xjjiang@pmo.ac.cn}\\
  \and
Shanghai Astronomical Observatory, Chinese Academy of Sciences, 80 Nandan Road,
Shanghai 200030, China\\
  \and 
School of Astronomy and Space Sciences, Nanjing University, Nanjing 210093, China\\
  \and
Key Laboratory of Modern Astronomy and Astrophysics (Nanjing
University), Ministry of Education, Nanjing 210093, China\\
  \and
Collaborative Innovation Center of Modern Astronomy and Space Exploration,
Nanjing 210093, China\\
}


   \date{Received~~2016 month day; accepted~~2016~~month day}

\abstract
{}
{We aim to systematically study the properties of the different transitions
of the dense molecular gas tracer \hcccn\ in galaxies.}
{We have conducted single-dish observations of \hcccn\ emission
lines towards a sample of nearby gas-rich galaxies. \hcccn ($J$=2-1) was observed
in 20 galaxies with Effelsberg 100-m telescope. \hcccn ($J$=24-23) was observed
in nine galaxies with the 10-m Submillimeter Telescope (SMT).}
{\hcccn\ 2-1 is detected in three galaxies: \object{IC 342},
\object{M 66} and \object{NGC 660} ($> 3\,\sigma$). 
\hcccn\ 24-23 is detected in three galaxies:
\object{IC 342}, \object{NGC 1068} and \object{IC 694}. This is the first
measurements of \hcccn\ 2-1 in a relatively large sample of external galaxies,
although the detection rate is
low. For the \hcccn\ 2-1 non-detections, upper limits (2\,$\sigma$) are derived
for each galaxy, and stacking the non-detections is attempted to recover
the weak signal of \hcccn. But the stacked spectrum does not show any
significant signs of \hcccn\ 2-1 emission. The results are also compared with
other transitions of \hcccn\ observed in galaxies. }
{The low detection rate of both transitions suggests low abundance of
\hcccn\ in galaxies, which is consistent with other observational
studies. The comparison between \hcccn\ and HCN or \hcop shows a
large diversity in the ratios between \hcccn\ and HCN or \hcop. More
observations are needed to interpret the behavior of \hcccn\ in different
types of galaxies.}
%For both $J$=2-1 and $J$=24-23 transitions \hcccn\ was not detected in the
%starburst galaxy M\,82, which supports the scenario that \hcccn\ is easily
%destroyed in the Photo-Dissociated Regions (PDRs) induced by the intense
%starburst of M\,82.

\titlerunning{\hcccn~ observations of Nearby Galaxies}  % title_head in odd pages
\authorrunning{Jiang, Wang, Gao \& Gu}            %author_head in even pages

   \maketitle

%
%_________________________________ sections below
%
\section{Introduction}           %% first-level sections will be auto-capitalized
\label{sect:intro}
Molecular lines play an essential role in our understanding of star formation
activity and galaxy evolution. With molecular lines of different species and
their different transitions, not only the chemical composition of the
interstellar medium can be investigated, but other important physical
parameters, such as temperature, pressure, density, and non-collisional pumping
mechanism can be derived as well (e.g., \citealt{Henkel1991};
\citealt{Evans:1999}; \citealt{Fukui:2010}; \citealt{Meier:2012};
\citealt{Meier:2014}).  New facilities providing wide band and highly sensitive
instruments are making weak line surveys and multi-species analysis feasible,
and the detections and measurements of a variety of species, are helping us to
reveal the gas components of galaxies, and how their abundances, densities,
ratios reflect the radiative properties of galaxies. Multi-species,
multi-transition molecular lines can be combined to diagnose the evolution stage
of galaxies \citep{Baan:2014}, because different species are sensitive to
different physical environments, such as photo dissociation regions (PDRs)
dominated by young massive stars,  X-rays dominated regions (XDRs) induced by
active galactic nucleus (AGNs), and shock waves by cloud-cloud collisions
\citep{Aladro2011, Greve2009, Aladro2011, Costagliola2011, Viti2014}.

%On the other hand, although  varied models have been proposed
%to connect molecular line emission with star formation (e.g. Krumholz et al. 2007,
%Meijerink et al. 2006), it is still hard to find a uniform scheme to illustrate how is
%molecular gas related to global environment  in different type of galaxies.

One of the interstellar species that benefits from the upgraded facilities is
cyanoacetylene (\hcccn). \hcccn\ was firstly detected in 1971 at 9.0977 GHz
($J$=2-1) in the Galactic star forming region Sgr B2 \citep{Turner1971}.
The critical density of \hcccn\ is comparable to the widely-used dense gas
tracer HCN and can also trace dense molecular gas around star forming sites.
%As a dense gas tracer ($\mu=3.6$ Debye, close to $\mu\sim$ 0.1 Debye for CO), 
\hcccn\ has been detected in many star formation regions in the Milky Way with
several transitions from centimeter to sub-millimeter (e.g.,
\citealt{Suzuki1992}). Due to the small rotational constant ($\sim$ 1/13 of
CO), there are many closely spaced rotational transitions of \hcccn\ (separated
by only 9.1 GHz) at centimeter and millimeter wavelengths, and its levels are
very sensitive to the changes in excitation \citep{Meier:2012}. This makes it
easier to conduct multi-transition observations of \hcccn\ lines than other
dense molecular gas tracers, and can help better understand the excitation
conditions of star forming regions. In contrast, the high-$J$ lines of other
dense molecular gas tracers such as HCN and HCO$^+$ are at very high
frequencies, thus it is difficult to observe them with ground-based telescopes.
Another advantage of using \hcccn\ lines is that \hcccn\ is very likely
optical thin even in low-$J$ transitions, due to the relatively low abundance
(\citealt{Irvine:1987}; \citealt{Lindberg2011}). And low opacity is important
for accurate estimate of dense molecular gas mass for the study of
relationship between dense molecular gas and star formation \citep{Gao2004a,
Gao2004b, Wang2011, Zhang:2014}.

There have been efforts to detect \hcccn\ in nearby galaxies, mainly in
millimeter band. Observations suggest that \hcccn\ is related to the warm,
dense star forming gas, and is easily dissociated by UV radiation
(\citealt{Henkel1988}; \citealt{Costagliola2011}; \citealt{Costagliola2011};
\citealt{Lindberg2011}; \citealt{Aladro2011}, \citealt{Aladro2015}). \hcccn\ was
found to be unusually luminous in \object{NGC 4418}, and it is attributed to its
high abundance (10$^{-7}$) as well as the intense radiation field in the dense
and warm gas at the center of \object{NGC 4418} \citep{Aalto2007,
Costagliola2010}. \citet{Meier2005, Meier:2012, Meier:2011, Meier:2014} presented high resolution observations of \hcccn\ ($J$=5-4,
10-9, 12-11 and 16-15) of a few very nearby galaxies, and gave detailed analysis
of the galactic structures and morphology that traced by \hcccn\ and other dense
gas tracers (HNC, HCN, CS, etc.). However, these results are still limited by
their sample size, and the chemical process of \hcccn\ is still unclear. Larger
samples are still necessary for analyzing the properties of \hcccn\ and how it
relate to other galactic parameters. In this paper we present the first
systematic survey of \hcccn\ ($J=2-1$) and \hcccn\ ($J=24-23$) in a relatively
large sample of nearby galaxies. And the results are compared with the
observations of \hcccn\ in other transitions.  The critical densities of
\hcccn\ $J=2-1$ and \hcccn\ $J=24-23$ are about $3\times 10^3\ \rm cm^{-3}$ and
$4 \times 10^6\ \rm cm^{-3}$, respectively. And the upper state energies ($E_u$)
of the two transitions are 1.3 K and 131 K, respectively
\citep{Costagliola2010}.

%%%%%%%%%%%%%%%%%%%%%%%%%% Table 1 %%%%%%%%%%%%%%%%%%%%%%%%%%%%%

\begin{table*}
\begin{center}
\caption[]{The source list of the 21 galaxies observed on \hcccn~ emission
using the Effelsberg 100m and the SMT telescope. The columns are:
(1) galaxy name,
(2) and (3) coordinates, (4) Heliocentric velocities,
(5) distances, (6) total infrared luminosities \citep[from][]{Sanders:2003},
(7) CO 1-0 linewidth (FWHM) of the galaxies, 
and (8) the telescope these galaxies were observed with. }
\begin{tabular}{lccrrrrl}
\hline
\hline\noalign{\smallskip}
Galaxy    & \multicolumn{2}{c}{RA  (J2000)   Dec} &  $V_{\rm Helio}$ & Distance 
&  log $L_{\rm IR}$ & $\Delta V_{\rm CO}$ & \ \ \ \ Telescope\\%&  \LIR$^{c}$&  Type$^{d}$ \\
		  &  h~ m~ s ~  & ~ \degr~~ \arcmin ~~ \arcsec  &
		   (km s$^{-1}$) & (Mpc) & ($L_\odot$) & (km s$^{-1}$) \\%($10^{10}$\Lsun)\\
\ \ \ \ \  (1) & (2) & (3) & (4) & (5) & (6) & (7) & \ \ \ \ \ \ \ \ (8) \\
  \hline\noalign{\smallskip}
NGC 520   & 01 24 34.9 & $+$03 47 30.0 & 2281 &   30.22 & 10.91 & 270  & Effelsberg \\ 
NGC 660   & 01 43 02.4 & $+$13 38 42.0 & 850  &   12.33 & 10.49 & 280  & Effelsberg \\
NGC 891   & 02 22 33.4 & $+$42 20 57.0 & 528  &   8.57  & 10.27 & 110  & Effelsberg \\ 
NGC 972   & 02 34 13.4 & $+$29 18 41.0 & 1543 &   20.65 & 10.67 & 220  & Effelsberg \\ 
NGC 1068  & 02 42 41.4 & $-$00 00 45.0 & 1137 &   13.7  & 11.27 & 280  & Effelsberg \& SMT \\ 
IC 342    & 03 46 48.5 & $+$68 05 46.0 &  31  &   4.60  & 10.17 & 72.8$^a$ & Effelsberg \& SMT \\
UGC 2855  & 03 48 20.7 & $+$70 07 58.0 & 1200 &   19.46 & 10.75 &\dots & Effelsberg \\ 
UGC 2866  & 03 50 14.9 & $+$70 05 40.9 & 1232 &   20.06 & 10.68 &\dots & Effelsberg \\ 
NGC 1569  & 04 30 49.0 & $+$64 50 53.0 &$-$104&   4.60  & 9.49  &  90  & Effelsberg \\ 
NGC 2146  & 06 18 39.8 & $+$78 21 25.0 & 882  &   16.47 & 11.07 & 320  & Effelsberg \& SMT \\ 
NGC 2403  & 07 36 51.3 & $+$65 36 29.9 & 161  &   3.22  & 9.19  &  90  & Effelsberg \\ 
M 82      & 09 55 53.1 & $+$69 40 41.0 & 187  &   3.63  & 10.77 & 150  & Effelsberg \& SMT \\ 
NGC 3079  & 10 01 57.8 & $+$55 40 47.0 & 1116 &   18.19 & 10.73 & 380  & Effelsberg \\ 
NGC 3310  & 10 38 45.9 & $+$53 30 12.0 & 993  &   19.81 & 10.61 & 140  & Effelsberg \\ 
M 66      & 11 20 15.0 & $+$12 59 30.0 & 727  &   10.04 & 10.38 & 180  & Effelsberg \\
IC 694    & 11 28 33.8 & $+$58 33 45.0 & 3120 &   47.74 & 11.63 & 250  & Effelsberg$^b$ \& SMT \\ 
NGC 3690  & 11 28 30.8 & $+$58 33 43.0 & 3120 &   47.74 & 11.32 & 260  & Effelsberg$^b$ \& SMT \\ 
Mrk 231   & 12 56 14.2 & $+$56 52 25.0 &12600 &   171.84& 12.51 & 167  & Effelsberg \\ 
Arp 220   & 15 34 57.1 & $+$23 30 10.0 & 5352 &   79.90 & 12.21 & 360  & Effelsberg \& SMT \\ 
NGC 6240  & 16 52 58.9 & $+$02 24 03.0 & 7160 &   103.86& 11.85 & 420  & \hspace{51.pt}~SMT \\ 
NGC 6946  & 20 34 52.6 & $+$60 09 12.0 & 53   &   5.32  & 10.16 & 130  & Effelsberg \& SMT \\ 
%IRAS18455+04 & 18 48 02.4 & $+$04 51 31.0 & 24&10.16 10.16 53     5.32  Effelsberg\\ 
  \hline\noalign{\smallskip}
\end{tabular}\label{Tab:source}
\tablefoot{
\tablefoottext{a}{for IC 342, the $\Delta V$ is of CO 2-1 from \citet{Gao2004a};
%\multicolumn{8}{l}{{\bf Notes.---} $^a$ for IC 342, the $\Delta V$ is of CO
%2-1 from \citet{Gao2004a};}\\
for other galaxies $\Delta V$ are of CO 1-0 from
\citet{Young1995}.}
\tablefoottext{b}{The Arp 299 system (IC 694 and NGC 3690)
was observed as a single pointing by the Effelsberg.}
}
\end{center}
\end{table*}

\section{Observations \& Data reduction}
\label{sect:Obs}

We select nearby infrared bright galaxies \citep{Sanders:2003} with IRAS 60
$\mu$m flux greater than 30 Jy and declination greater than $-21\degr$ to do
this survey. It is not a complete but representative sample of infrared bright
galaxies.  The sample consists of 21 galaxies. Note that due to the different
beam size of the two telescopes we used, the merger
Arp 299 (\object{IC 694} and \object{NGC 3690}) were observed as a single
pointing by Effelsberg 100-m, while the two galaxies were observed separately by
the SMT 10-m.

\subsection{\hcccn\ 2-1 observations with the Effelsberg 100-m}\label{sect:obs_21}
\hcccn\ ($J=2-1$) ($\nu_{\rm rest} =$ 18.196 GHz) of 20 galaxies was observed
with Effelsberg 100-m telescope in 2010. The Half Power Beam Width (HPBW) is
46.5$''$ at 18 GHz for the 100-m telescope. We used the 1.9 cm band receiver,
500 MHz bandwidth with 16384 channels correlator setup, which provided $\sim$
8300 \kms\ velocity coverage and $\sim$ 0.5 \kms\ velocity resolution
during the observations. Position-switching mode with beam-throws of about $\pm
2'$ was used. Pointing and focus were checked about every two hours. The typical
system temperature of the Effelsberg observations was about 46 K. The
on-source time for each galaxy is about 14--47 minutes.  The weather during the
observations is not ideal, and the baselines of many sources are affected and
induced artificial features which are hard to remove.

\subsection{\hcccn\ 24-23 observations with the SMT 10-m}\label{sect:obs_2423}
\hcccn\ ($J=24-23$) ($\nu_{\rm rest} =$ 218.324 GHz) of nine galaxies was
observed in 2009 with the SMT 10-m telescope. The HPBW is about 33$''$ at
$\sim$218 GHz for SMT, and a single pointing was used
for each galaxy toward their central positions. We used the ALMA Sideband
Separating
Receiver and the Acousto-Optical-Spectrometers (AOS), which have dual
polarization, 970 MHz ($\sim$ 1300 \kms) bandwidth and 934kHz channel spacing.
Observations were carried out with the beam-switching mode with a
chop throw of $2'$ in azimuth (AZ) and a chopping frequency of 2.2 Hz. Pointing
and focus were checked about every two hours by measuring nearby QSOs with
strong millimeter continuum emission. The typical system temperature at 218 GHz
was less than 300 K, and the on-source time for each galaxy was $\sim$
60--168 minutes. 

\subsection{Data Reduction}\label{sect:reduction} 
The basic parameters of our
sample galaxies are listed in Table\ref{Tab:source}.  The data were reduced
with the CLASS program of the
GILDAS\footnote{\url{http://iram.fr/IRAMFR/GILDAS/}} package. First, we checked
each spectrum and discarded the spectra with unstable baseline.  Most of the
Effelsberg spectra do not have flat baselines, but over several hundred \kms\
near the line the
baselines can still be fixed. In the SMT spectra the image signal of strong CO
2-1 in the upper sideband affect the baseline of the lower sideband and for
\object{M\,82} and \object{Arp\,220} the \hcccn\ 24-23 is contaminated. But for
other galaxies the image CO
line does not affect the \hcccn\ line.  Then we combined spectra with both
polarizations of the same source into one spectrum.  Depending on the quality
of the spectral baselines, a first-order or second-order fitting was used
to subtract baselines from all averaged spectra. The identifications of the
transition frequencies of \hcccn\ have made use of the NIST database
{\it Recommended Rest Frequencies for Observed Interstellar Molecular Microwave
Transitions}\footnote{\url{http://www.nist.gov/pml/data/micro/index.cfm}}. 

To reduce the noise level, the spectra are smoothed to velocity
resolutions $\sim 20-40$ \kms.  The velocity-integrated intensities of
the \hcccn\ line are derived from the Gaussian fit of the spectra, 
or integrated over a defined window if the line profiles significantly deviate
from a Gaussian. The intensities are
calculated using $ I = \int  T_{\rm mb}{\rm d}v$,
where $T_{\rm mb}$ is the main beam brightness temperature. Molecular line
intensity in antenna temperature ($T_A^{*}$) is converted to main beam
temperature $T_{\rm mb}$ via $T_{\rm mb} = T_A^{*} /{\rm MBE}$, with the main
beam efficiency MBE = 53\% at 18 GHz for Effelsberg telescope, and 70\%
at 218 GHz for SMT during the observations. The flux density is then
derived from $T_{\rm mb}$, using $S/T_{\rm mb}$ = 0.59 Jy/K for the Effelsberg 
telescope, and 24.6 Jy/K for the SMT. 

%\bf calibration: T$_{mb}$ = QSO$_{standard}$(Jy) / (QSO$_{obs}$(K)/Sensitivity [K/Jy]) $\times$ T$_A^*$/MBE

\section{Results \& Discussion}\label{sect:results}
%\subsection{\hcccn~$J$=2-1}
The spectral measurements and estimated intensities of the \hcccn\ lines, including
RMS noise and on-source time, are listed in Table\ref{Tab:results} (\hcccn\
2-1) and Table\ref{Tab:results2} (\hcccn\ 24-23).

\subsection{\hcccn\ 2-1} \label{sect:hcccn21}
\begin{figure*}
 %  \centering
   \includegraphics[scale=.25, angle=0]{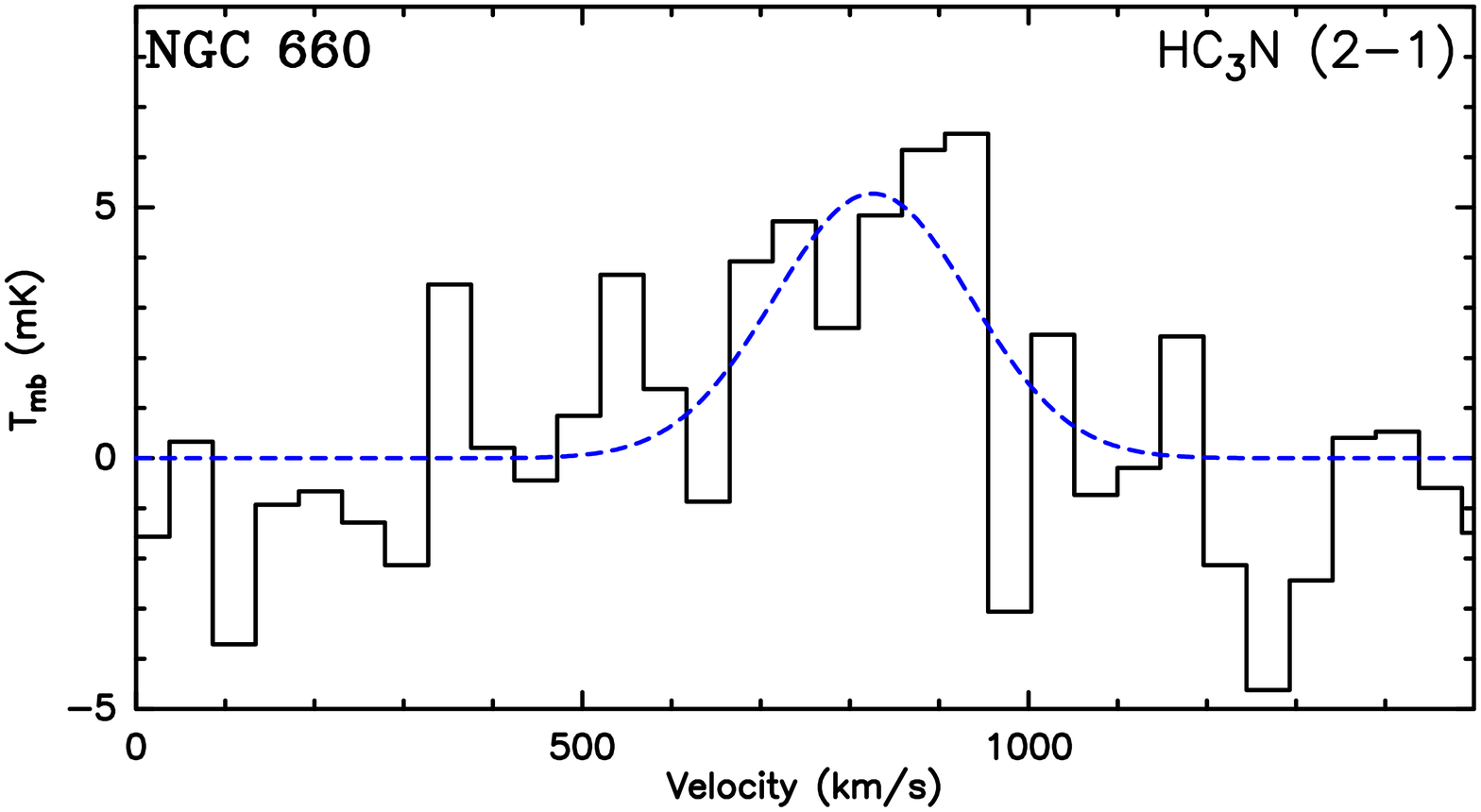}
   \includegraphics[scale=.25, angle=0]{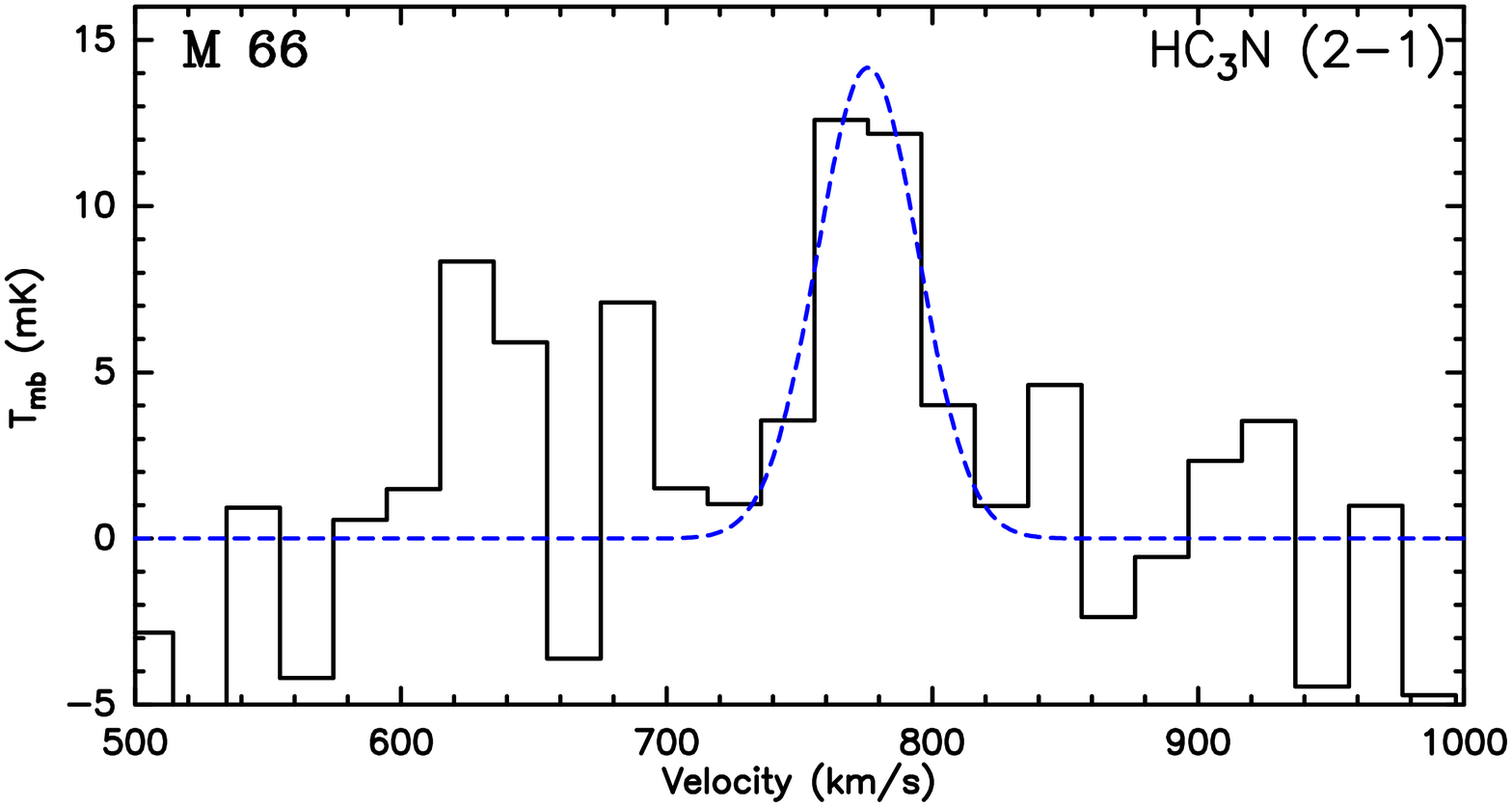}\\
   \includegraphics[scale=.25, angle=0]{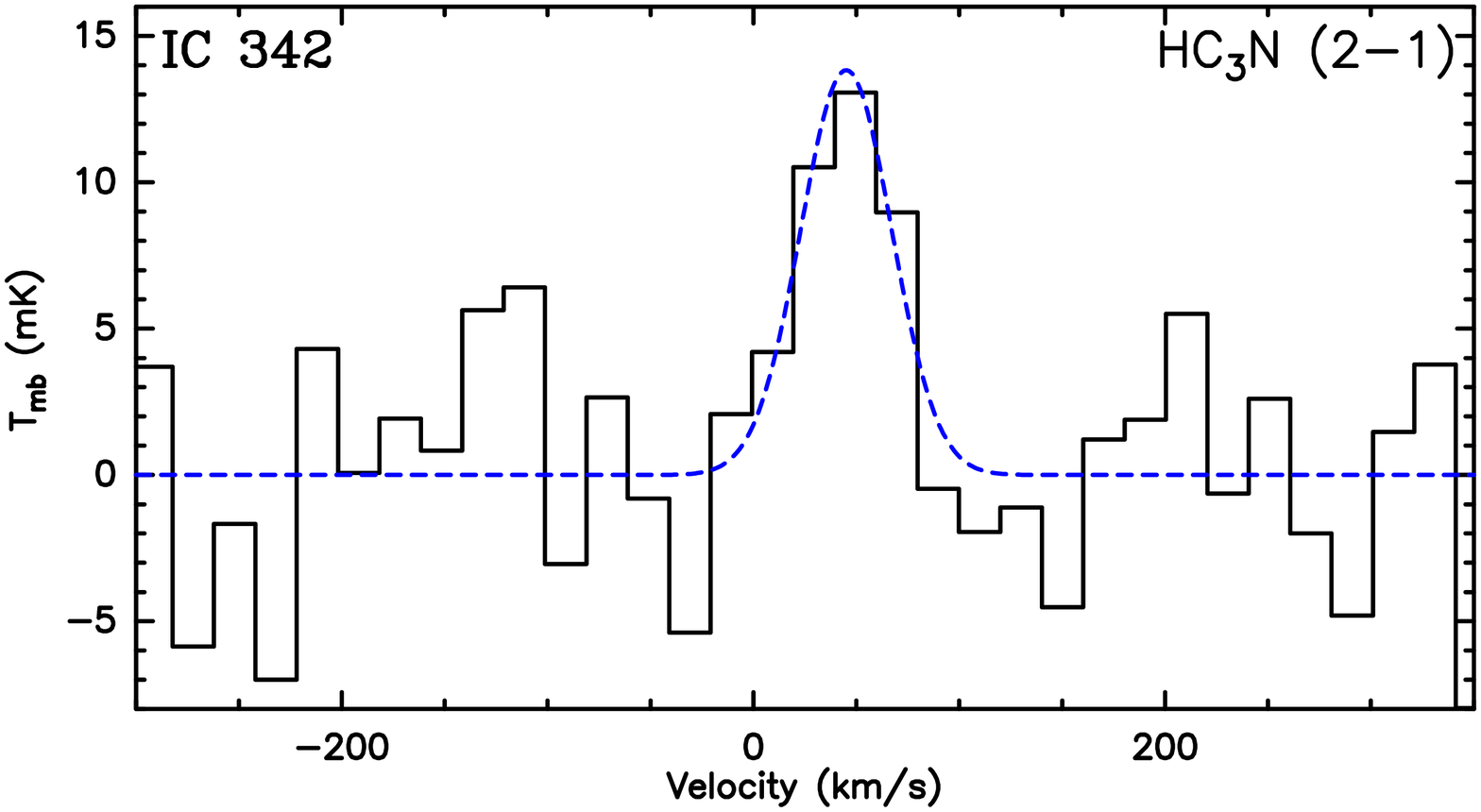}
   \includegraphics[scale=.25, angle=0]{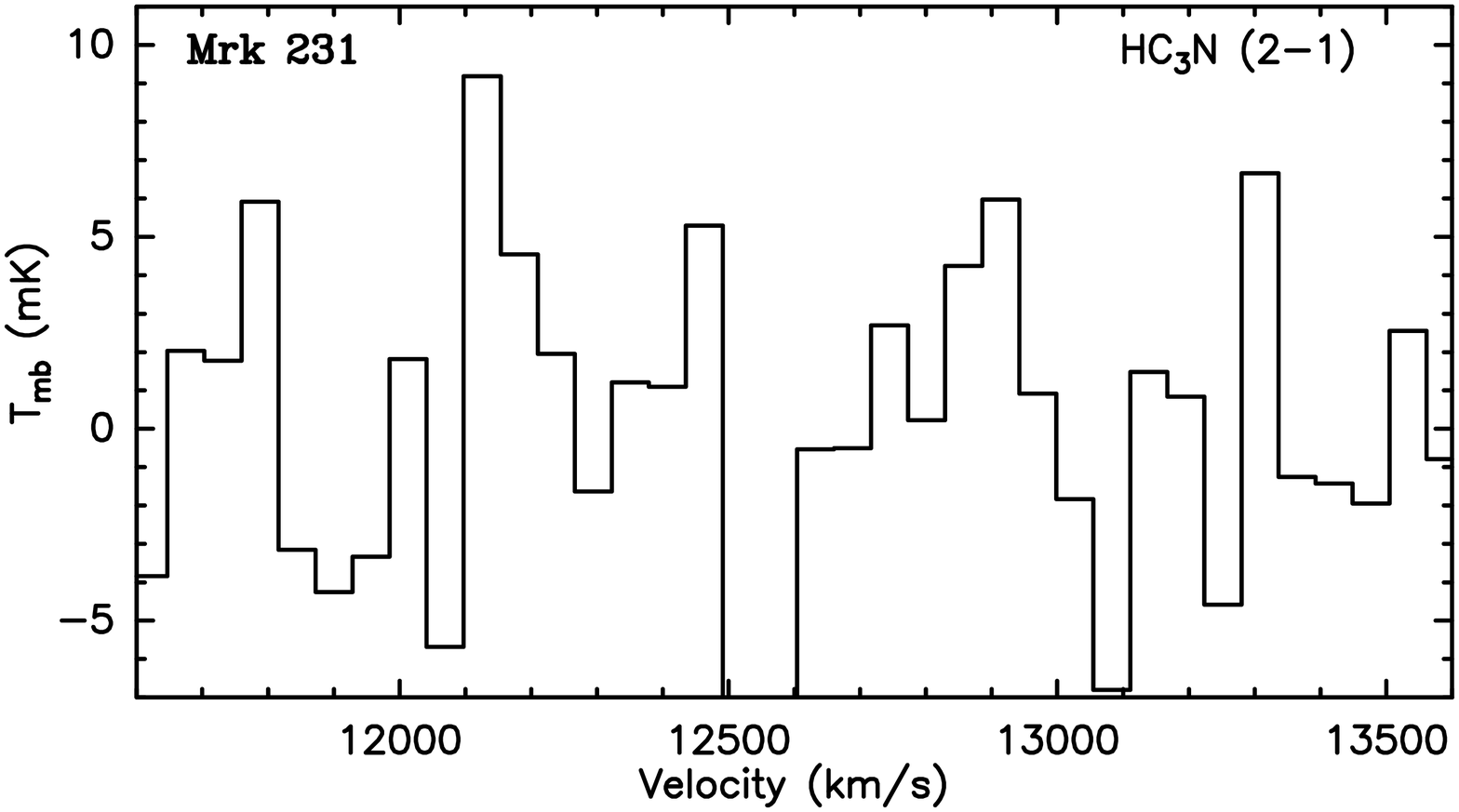}\\
   \includegraphics[scale=.25, angle=0]{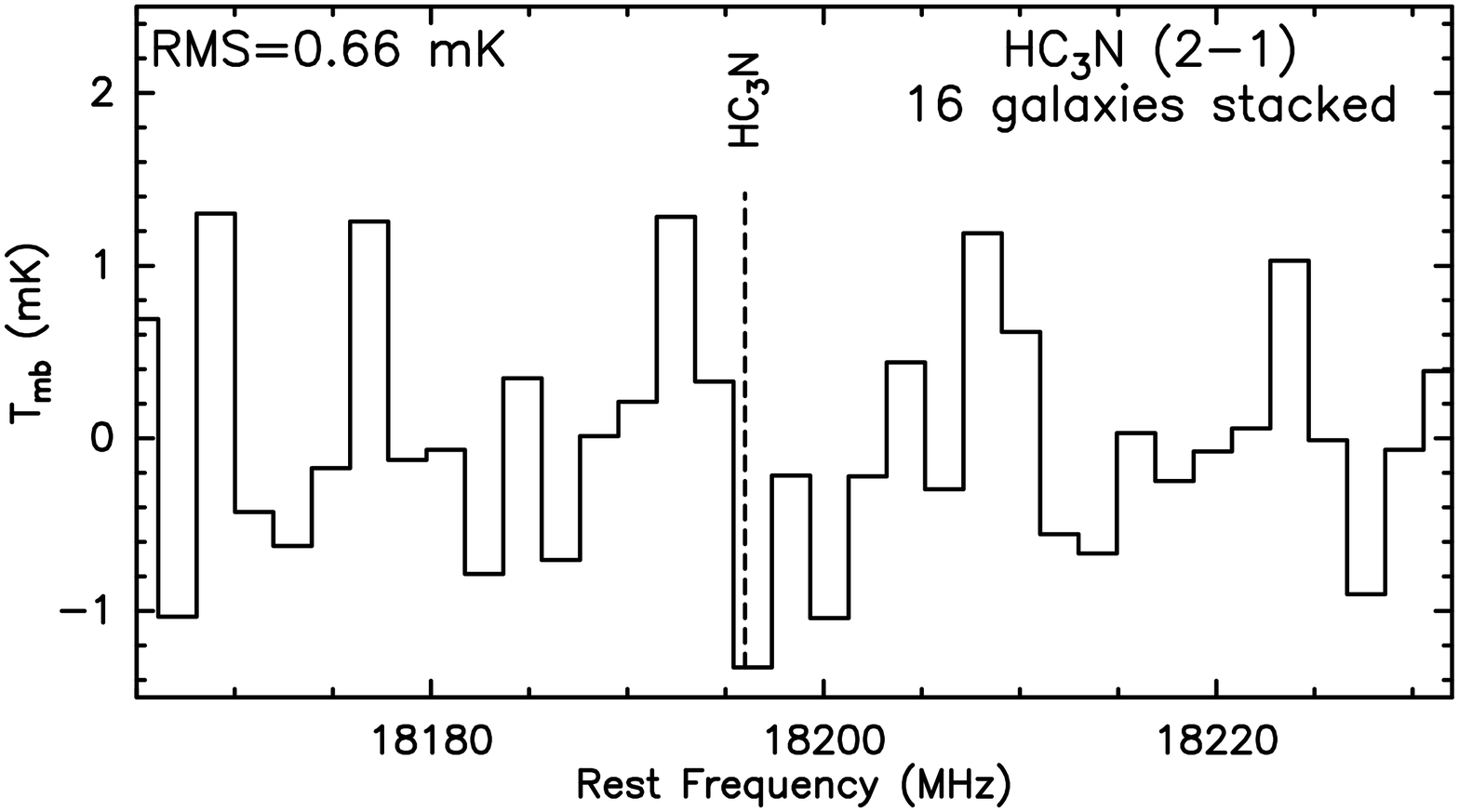}
   \includegraphics[scale=.25, angle=0]{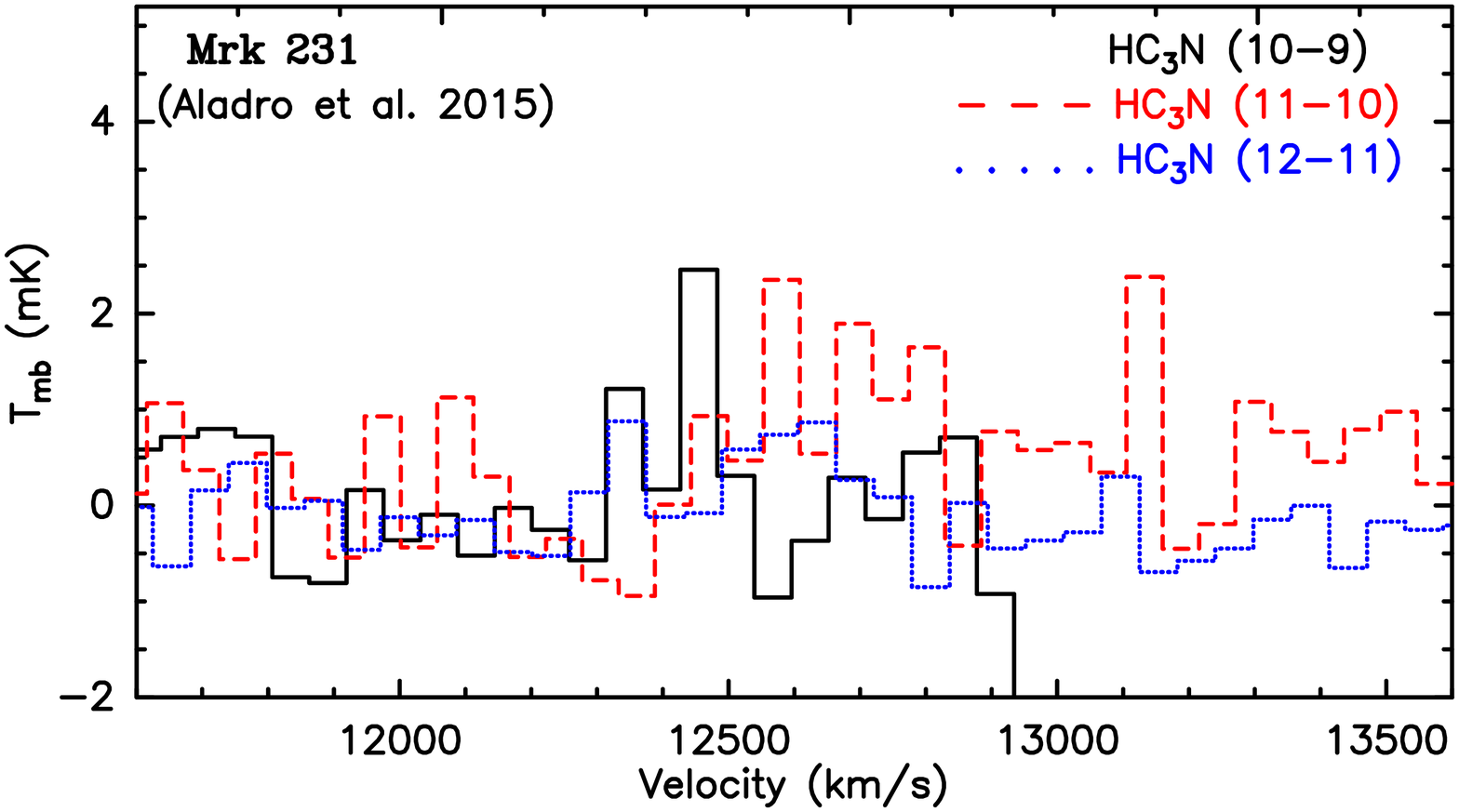}\\
   \caption{Spectra
   of the detected \hcccn($J$=2-1) in NGC 660, M 66, IC 342 by the
   Effelsberg. At the bottom row also shows the stacked spectra and the
   spectra of Mrk 231 from \citet{Aladro2015}. Blue dashed lines are the
   Gaussian fit of the \hcccn\ 2-1 lines. Temperature scale is $T_{\rm mb}$
   in mK. A colour version of this figure is available in the online journal.} 
   \label{Fig:2_1}
\end{figure*}

Among the 20 galaxies observed by Effelsberg 100-m telescope, \hcccn\ 2-1 is
detected in three galaxies: \object{IC 342}, \object{NGC 660} and \object{M 66} (See
Figure\ref{Fig:2_1}). This is the first report of \hcccn\ 2-1 detections in
external galaxies, although limited by the SNR (Signal to Noise Ratio) the detection rate is low.  

\paragraph{\textit{\object{IC 342}}:} IC 342
has the strongest peak intensity ($T_{\rm mb}\sim$ 14 mK) of \hcccn\ 2-1 in the
sample, which
is about twice the strength as the \hcccn(9-8) line of IC 342 detected by IRAM
30-m telescope \citep{Aladro2011}, while the line width (FWHM $\sim$ 60 \kms)
is similar to their result. 

\paragraph{\textit{\object{NGC 660}}:} The detected \hcccn\ 2-1 in \object{NGC
660} has a similar
linewidth (FWHM $\sim$ 294.7 \kms) to CO 1-0 ($\sim$ 280 \kms). While the
\hcccn\ survey by \citet{Lindberg2011} did not observe \object{NGC 660}, its 10-9
and 12-11 transitions were not detected
by \citet{Costagliola2011}. This difference in the detection of \hcccn\ lines may
imply that there is little warm and dense gas content in \object{NGC 660}, thus the
high-$J$ \hcccn\ lines can not be excited.  

\paragraph{\textit{\object{M 66}}:} In \object{M 66} \hcccn\ 2-1 is only detected on about 2
$\sigma$ level, but this is the first tentative detection of \hcccn\ in \object{M 66}. It
was not observed by \citet{Costagliola2011} nor \citet{Lindberg2011}.

\paragraph{\textit{non-detections}:} Due to the poor quality (and probably
insufficient integration time) of the \hcccn\ 2-1 data, 16 out of 19 galaxies
were not detected. Assuming their linewidth is approximate to CO 1-0 linewidth
(FWHM, from \citealt{Young1995}), we derive upper limits of the integrated
intensity for each galaxy (2 $\sigma$, where $\sigma =$ RMS $\sqrt{\delta V
\cdot \Delta V}$) and show them in Table\ref{Tab:results}. Note that the
linewidth of \hcccn\ is likely narrower than that of CO, and such assumption
might overeistimate the upper limits of integrated intensity, thus is only
a rough estimate. The upper limits are
in the range of $\sim$ 0.3 -- 1.2 K \kms. For those non-detection galaxies, We
also
stack their spectra together, weighted by the RMS level of each galaxy, to exam
if a cumulated signal can be obtained (see Figurer \ref{Fig:2_1}). Although the
RMS of the the stacking \hcccn\ 2-1 spectrum is reduced down to 0.66 mK, we do
not see any signs of emission (at a resolution of 30 \kms). Since these galaxies
have similar linewidth (100--400 \kms), we can estimate the stacked upper limit
assuming a linewidth of 200 \kms, based on the RMS (0.66 mK) of the stacked
spectrum. Thus the 2 $\sigma$ upper limit of these galaxies is about 0.26 K
\kms. To eliminate the possible effect induced by different linewidths of
galaxies, we also tried to group the non-detection galaxies based on their CO
linewidth. Galaxies with CO FWHM (Table \ref{Tab:source}) wider than 200 \kms\
are stacked as one group, and other galaxies are stacked as another group.
Neither group shows any signs of emission.

%\subsection{\hcccn~$J$=24-23}
\subsection{\hcccn\ 24-23} 
\begin{figure*}
   \includegraphics[scale=.25, angle=0]{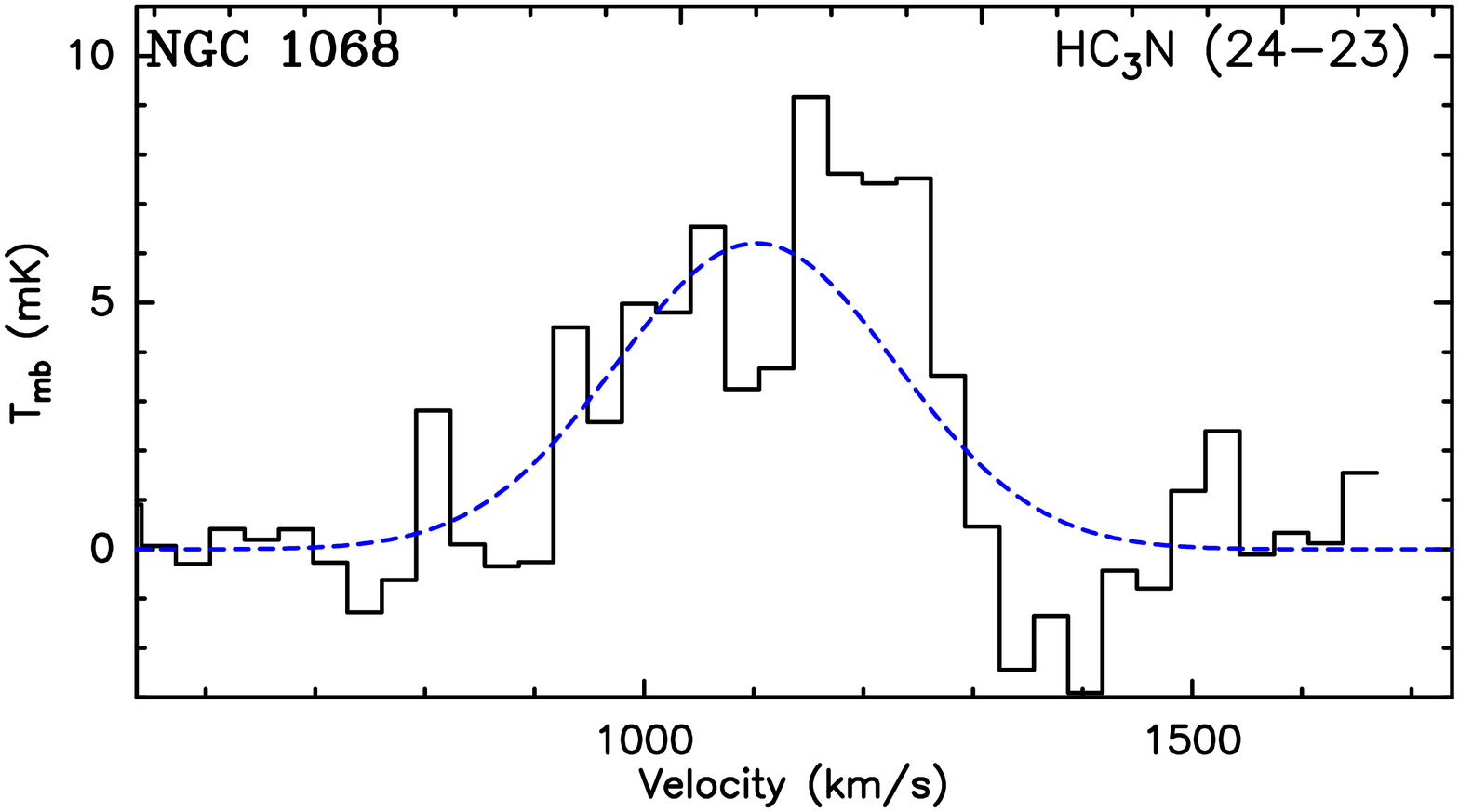}
   \includegraphics[scale=.25, angle=0]{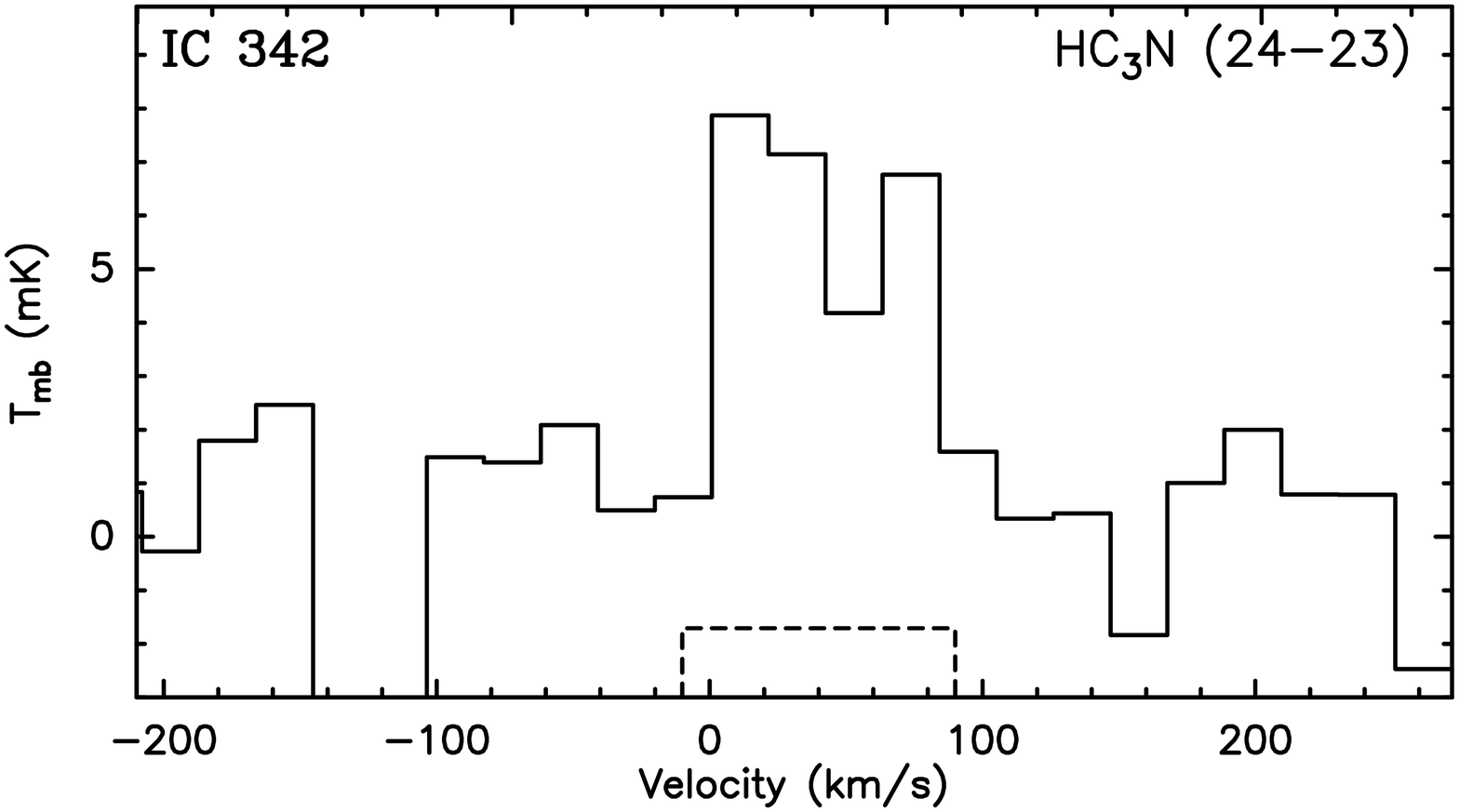}\\
   \includegraphics[scale=.25, angle=0]{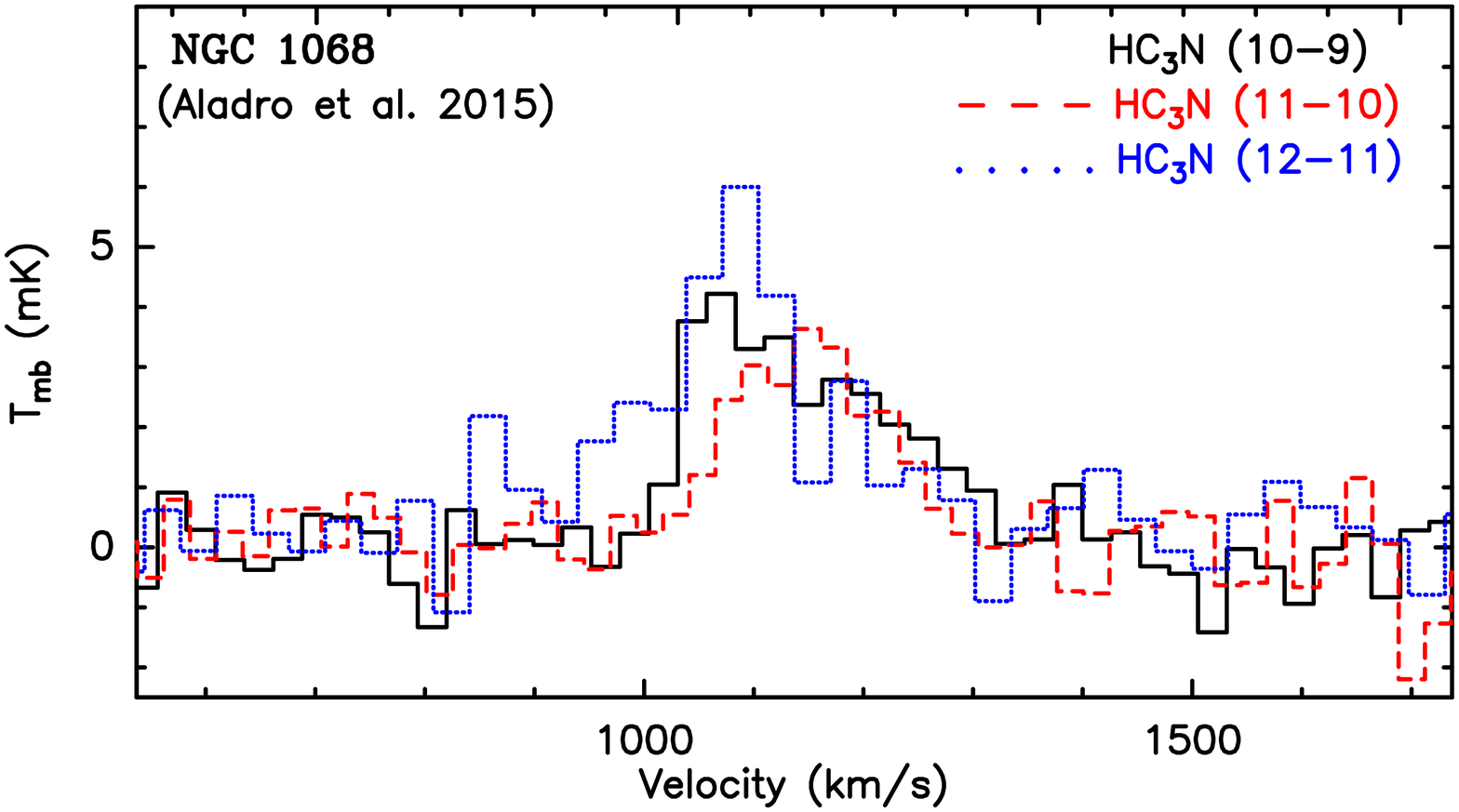}
   \includegraphics[scale=.25, angle=0]{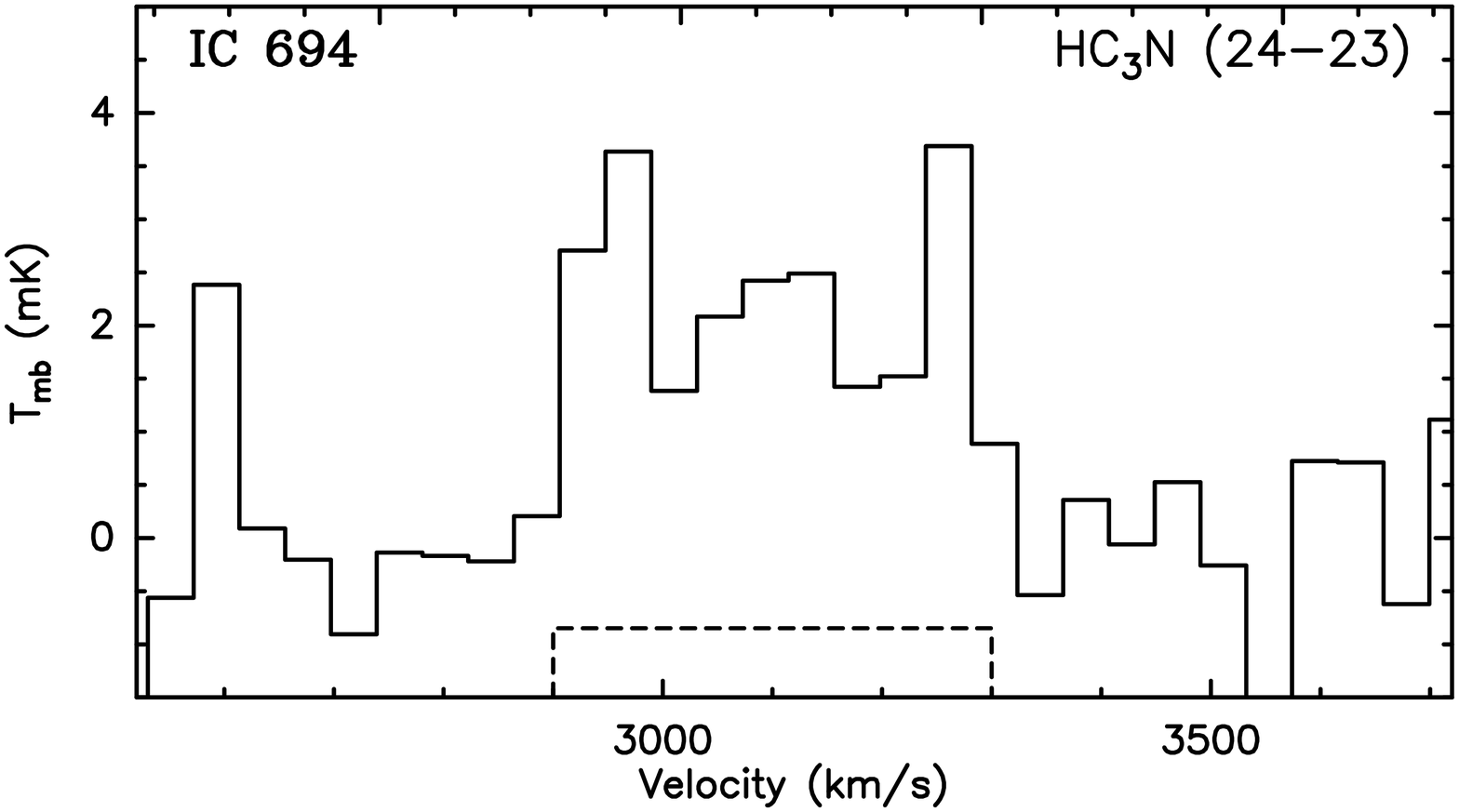}
   \caption{Spectra of detected \hcccn($J$=24-23) in \object{NGC 1068},
   \object{IC 342} and \object{IC 694}
   by the SMT. At the bottom left also shows the spectra of \object{NGC 1068} from
\citet{Aladro2015} for comparison. Blue dashed line is the Gaussian fit. Note that in IC 694 it is difficult to distinguish the 
\hcccn\ emission from the possibly blended \hhco\ lines. Temperature scale is $T_{\rm mb}$
in mK. A colour version of this figure is available in the online journal.} 
\label{Fig:24_23} 
\end{figure*}

Among the nine galaxies observed by SMT,
\hcccn($J$=24-23) is detected in three galaxies: \object{IC 342},  \object{NGC
1068} and  \object{IC 694}
(Figure\ref{Fig:24_23}).

\paragraph{\textit{ \object{IC 342}}:} \hcccn\ 24-23 of IC 342 was previously detected
and measured by \cite{Aladro2011}, and our observation obtains consistent
result, although comparing to their observation we do not detect \hhco\
simultaneously. In our observations, IC 342 is the only galaxy detected in
both 2-1 and 24-23 transitions. The line center and width of the two transitions 
are similar, considering observational uncertainties. This might imply that 
the two transitions have similar emitting area. And the ratio between the
integrated intensities of \hcccn\ 24-23/\hcccn\ 2-1 is about 0.6.

%The \hcccn\ 10-9/HCN(1-0) ratio in IC 342 was 0.28$\pm$0.08 derived by
%\citet{Lindberg2011}, and based on our data, \hcccn\ 2-1/HCN(1-0)
%$= 0.15\pm$0.02, and . This is about half that of \hcccn\ 10-9/HCN(1-0), but more
%close to the ratio of other galaxies (e.g. I7208, see Table. 7 in
%\citealt{Lindberg2011}).

\paragraph{\textit{ \object{NGC 1068}}:} In NGC 1068, the integrated intensity of
\hcccn\ 24-23 is about 2.0 K \kms\ (in $T_{\rm mb}$), which is stronger than
that of \hcccn\ 10-9 ($\sim$1.1 K \kms) reported by \cite{Costagliola2011}. It
may imply that there is sufficient warm and dense gas, which is able to excite
the high transition \hcccn\ 24-23 line.  Besides, it could also be affected by
the strong AGN signature of this galaxy \citep{Wang:2014, Tsai:2012}.

\paragraph{\textit{ \object{IC 694}}:} Previous observation only obtained upper limits
of \hcccn\ 12-11 For IC 694 \citep{Lindberg2011}. In our observations, a
tentative detection in IC 694 ($> 2
\sigma$) is obtained. The line profile of IC 694 obviously deviates from a
Gaussian, so we derive the \hcccn\ intensity by integrating the line within a
window of 400 \kms\ width (Table \ref{Tab:results2}).

We note that, in NGC 1068 and IC 694, \hcccn\ 24-23 is possibly blended
with H$_2$CO 3(0,3) -- 2(0,2) emission ($f_\nu=$218.22219 GHz). The upper state
energy of this is para-H$_2$CO line is about 10.5 K, which is likely to be
excited in these cases. The H$_2$CO line is shifted by 141.1 \kms or -102 MHz
from the \hcccn\ 24-23 line, and it is unclear that how much intensity of
\hcccn\ 24-23 in NGC 1068 and IC 694 is contributed by H$_2$CO (see
Figure\ref{Fig:24_23}). We still lack enough data to disentangle this issue, and
can only compare with other observations. For example, in the observation of M
82 by \citet{Ginard2015}, they showed that near the frequency of 145 GHz,
H$_2$CO 2(0,1) -- 1(0,1) is as strong as \hcccn\ 16-15.  H$_2$CO is not detected
in M 82 in 3mm band \citep{Aladro2015}.  In the observations toward NGC 4418 by
\citet{Aalto2007}, they showed that \hcccn\ 16-15 is blended with H$_2$CO ,and
H$_2$CO may contribute 20\% of the total integrated line intensity.

\paragraph{\textit{non-detections}:} The spectra of M 82 and Arp 220 are
seriously contaminated by the image signal of CO 2-1 from the upper side-band
($\nu$ = 230 GHz), which is strong and wide hence difficult to remove. As a
consequence we could not extract the spectrum of \hcccn\ properly. We treat the
\hcccn\ 24-23 in M 82 and Arp 220 as non-detections, and their 2$\sigma$ upper
limits are also only indicative. Although not contaminated by adjacent CO image
signal, \hcccn\ 24-23 was not detected in NGC 2146, NGC 6946, NGC 3690 and
NGC 6240. For these non-detection we present 2$\sigma$ upper limit of the
integrated intensity of \hcccn\ 24-23 in Table\ref{Tab:results2}. Only four
galaxies are not contaminated by CO image signal, thus no stacking is
implemented for their \hcccn\ 24-23 spectra. 

%%%%%%%%%%%%%%%%%%%%%%%%%%%%%%%%%%%%%%%%%%%%%%%%%%%%%%%%%%%%%%
%%     Examples for figures using graphicx for LaTeX 2e
%%               -- our recommended way for embodying graphics
%%%%%%%%%%%%%%%%%%%%%%%%%%%%%%%%%%%%%%%%%%%%%%%%%%%%%%%%%%%%%%
%
%      A figure as large as the width of the column
%-------------------------------------------------------------

%%%%%%%%%%%%%%%%%%%%%%%%%%%%%%%%%%%%%%%%%%%%%%%%%%%%%%%%%%%%%%%%%%
                       %  Table 2 % 

\begin{table*}
\begin{center}
\caption[]{\hcccn\ 2-1\ spectral measurements. 
All the temperature scales are $T_{\rm mb}$. The columns are: 
(1) galaxy name,
(2) on-source time for each galaxy,
(3) RMS noise of the smoothed spectrum,
(4) velocity resolution of the smoothed spectrum, 
(5) linewidth (FWHM) of the Gaussian fit of the line (if available),
(6) \hcccn\ emission line center, 
(7) Integrated intensity (and errors) of \hcccn\ 2-1 emission. For those 
non-detections, 2 $\sigma$ upper limits are presented (see text in Section
\ref{sect:hcccn21}),
(8) Integrated flux density, and 
(9) flux density ratio between \hcccn\ 2-1 and HCN 1-0.
}\label{Tab:results}
%%Please Capitalize the First Letter of Each Notional Word in table's caption
\begin{tabular}{ccccccccc}
\hline
  \hline\noalign{\smallskip}
Source  &  On-time  & RMS & $\delta V$  &  $\Delta V$  & $V_0$ &
$I(\rm HC_3N)$  & $S(\rm HC_3N)$ & $\frac{\rm HC_3N\ 2-1}{\rm HCN\ 1-0}$  \\
%\cline{4-6} \cline{7-10}
  & (min) & (mK) & (km s$^{-1}$) 
  &(km s$^{-1}$) & (km s$^{-1}$) & (K km s$^{-1}$)  &  (Jy km s$^{-1}$)  \\  
 (1) & (2) & (3) & (4) & (5) & (6) & (7)  & (8)  & (9)\\
%%Type\tablenotemark{b}} \\
\hline\noalign{\smallskip}
%& \multicolumn{5}{c}{\hcccn~$J$=2-1} \\
%\cline{2-6}\noalign{\smallskip}
NGC 660   &  28  & 4.2  &  32.2  & 260.7 (90.6) & 825 (36)& 1.47 (0.40)   & 0.86 (0.24)   & $\sim$ 0.034$^a$\\
IC 342    &  20  & 6.5  &  20.1  & 52.0 (12.9)  & 45 (6)  & 0.77 (0.18)   & 0.45 (0.11)   & $\sim$ 0.005$^b$\\
M 66      &  18  & 7.0  &  20.1  & 44.8 (17.7)  & 775 (7) & 0.68 (0.21)   & 0.4 (0.12)   & $\sim$ 0.114$^c$\\
NGC 520   &  14  &  5.7 & \dots  &   \dots   &    \dots   & $<$ 1.02      & $<$ 0.6 	& $<$ 0.087$^d$ \\  
NGC 891   &  30  &  2.3	& \dots  &   \dots   &    \dots   & $<$ 0.26      & $<$ 0.15  & $<$ 0.033$^e$  \\ 	
NGC 972   &  33  &  3.8	& \dots  &   \dots   &    \dots   & $<$ 0.61      & $<$ 0.36  & \dots    \\ 	
NGC 1068  &  25  &  4.0	& \dots  &   \dots   &    \dots   & $<$ 0.73      & $<$ 0.43  & $<$ 0.013$^f$ \\ 	
UGC 2855  &  23  &  4.9	& \dots  &   \dots   &    \dots   & $<$ 0.76$^*$  & $<$ 0.45  &  \dots   \\ 	
UGC 2866  &  22  &  3.6	& \dots  &   \dots   &    \dots   & $<$ 0.56$^*$  & $<$ 0.33  &  \dots   \\ 	
NGC 1569  &  33  &  3.0	& \dots  &   \dots   &    \dots   & $<$ 0.31      & $<$ 0.18  &  \dots   \\ 	
NGC 2146  &  24  &  3.4	& \dots  &   \dots   &    \dots   & $<$ 0.67      & $<$ 0.39  & $<$ 0.021$^e$ \\ 	
NGC 2403  &  30  &  3.2	& \dots  &   \dots   &    \dots   & $<$ 0.33      & $<$ 0.19  &  \dots   \\ 	
M 82      &  26  &  5.5	& \dots  &   \dots   &    \dots   & $<$ 0.73      & $<$ 0.43  & $<$ 0.011$^e$ \\ 	
NGC 3079  &  19  &  5.5 & \dots  &   \dots   &    \dots   & $<$ 1.17      & $<$ 0.69	& $<$ 0.225$^e$ \\  
NGC 3310  &  36  &  2.3	& \dots  &   \dots   &    \dots   & $<$ 0.29      & $<$ 0.17  &  \dots   \\ 	
IC 694+NGC3690 &47& 4.2	& \dots  &   \dots   &    \dots   & $<$ 0.72      & $<$ 0.42  & $<$ 0.158$^g$ \\ 	
%NGC 3690  & \dots& \dots& \dots  &   \dots   &&   \dots     \dots        & %\dots      %\d\\ 
Mrk 231   &  44  &  6.4	& \dots  &   \dots   &    \dots   & $<$ 0.91      & $<$ 0.54  & $<$ 0.191$^g$ \\ 	
Arp 220   &  26  &  4.5	& \dots  &   \dots   &    \dots   & $<$ 0.94      & $<$ 0.55  & $<$ 0.069$^h$ \\ 	
NGC 6946  &  28  &  5.1	& \dots  &   \dots   &    \dots   & $<$ 0.64      & $<$ 0.38  & $<$ 0.018$^e$         \\ 	
\hline\noalign{\smallskip}
\end{tabular}
\tablefoot{
\tablefoottext{*}{ The upper limits 
of the two galaxies are derived assuming a 200 \kms\ linewidth.}\\
HCN 1-0 data from: 
\tablefoottext{a}{\citet{Baan2008}; }
\tablefoottext{b}{\citet{Nguyen1992};}
\tablefoottext{c}{\citet{Krips2008};}
\tablefoottext{d}{\citet{Solomon1992}; } 
\tablefoottext{e}{\citet{Gao2004a}; }
\tablefoottext{f}{\citet{Aladro2015};}  
\tablefoottext{g}{\citet{Jiang2011}; }
\tablefoottext{h}{\citet{Wang2016}.}
}\\
%\multicolumn{4}{l}{ $^b$ CO width from \citet{Greve2009}; }\\
%\multicolumn{4}{l}{$^{(1)}$ CO width from \citet{Young1995}.}
\end{center}
\end{table*}

\begin{table*}
\begin{center}
\caption[]{\hcccn\ 24-23\ spectral measurements. Columns are 
the same as Table \ref{Tab:results}.
} \label{Tab:results2}
\begin{tabular}{ccccccccc}
\hline
  \hline\noalign{\smallskip}
Source  &  On-time  & RMS & $\delta V$&  $\Delta V$  & $V_0$ &
$I(\rm HC_3N)$ &  $S(\rm HC_3N)$ & $\frac{\rm HC_3N\ 24-23}{\rm HCN\ 1-0}$  \\
%\cline{4-6} \cline{7-10}
  & (min) & (mK) & (km s$^{-1}$) & (km s$^{-1}$)
  &(km s$^{-1}$) & (K km s$^{-1}$) & (Jy km s$^{-1}$) \\  
 (1) & (2) & (3) & (4) & (5) & (6) &(7) &(8) &(9)  \\
%%Type\tablenotemark{b}} \\
\hline\noalign{\smallskip}
NGC 1068& 121 &   1.2  & 39.1 & 257.3 (24.0)$^a$  & 1102 (13) & 2.03 (0.18) & 49.9 (4.4)& $\sim$ 0.59 \\
IC 342  & 132 &   2.7  & 20.9 &    100$^b$        &  43       & 0.47 (0.12) & 11.6 (3.0)& $\sim$ 0.12          \\        
IC 694  & 127 &   0.87 & 41.7 &    400$^b$        & 3095      & 0.90 (0.11) & 22.1 (2.7)& $\sim$ 2.76          \\       
NGC 2146& 115 &   1.01 & 20.9 &  320$^{c}$        &  \dots    & $<$ 0.17    & 4.2& $<$  0.26      \\   
M 82    &  60 &   3.0  & 20.9 &  150$^{c}$        &  \dots    & $<$ 0.34    & 8.4& $<$  0.26      \\   
NGC 3690& 139 &   0.73 & 42.0 &    260$^c$        &  \dots    & $<$ 0.15    & 3.7& $<$  1.27     \\        
ARP 220 & 162 &   0.75 & 39.1 &    420$^c$        &  \dots    & $<$ 0.19    & 4.7& $<$  0.48     \\        
NGC 6240&  97 &   0.92 & 20.1 &  420$^{c}$        &  \dots    & $<$ 0.17    & 4.2& $<$  1.15      \\   
NGC 6946& 168 &   1.76 & 21.0 &  130$^{c}$        &  \dots    & $<$ 0.18    & 4.4& $<$  0.26      \\   
\hline\noalign{\smallskip}
\end{tabular}
\tablefoot{
\tablefoottext{a}{ FWHM from the Gaussian fitting and error.}
\tablefoottext{b}{ line window (full width) used to derived the integrated intensity.}
\tablefoottext{c}{ CO width from \citet{Young1995} that are used to derived the upper limits.}
}
\end{center}
\end{table*}
\subsection{Discussion: \hcccn\ in galaxies}

The HPBW of SMT and Effelsberg observations are 33$''$ and 46$''$,
respectively, which should be able to cover the bulk of the sample galaxies, especially
the galaxy center. Thus our observations should be able to cover the region
where the majority of dense gas resides. However, with single-dish observations
we can not constrain the emission size of either \hcccn\ 2-1 or \hcccn\ 24-23,
and can not easily estimate the filling factors. Along with the large
uncertainty of the emission intensities measurements, it is difficult to estimate
the brightness temperature of the sample.

To better understand the excitation environment of \hcccn, the effect of
free-free and synchrotron emission near 18 GHz should be also taken into
account, as they are more prominent than that in millimeter band that is
dominated by dust thermal emission. We detect \hcccn\ 2-1 lines in emission and
not in absorption, and this may be due to the fact that the beam filling factor
of the \hcccn\ gas is higher than the radio continuum.  In the high resolution
radio observations towards a few nearby galaxies \citep{Tsai2006}, it is found
that compact radio sources contribute  20\% -- 30\% of the total 2 cm (15 GHz)
emission from the central kiloparsec of these galaxies. In contrast, the
distribution of gas with moderate critical density such as \hcccn\ 2-1 is likely
more diffuse.

Comparing to other dense molecular gas tracers such as the popular HCN and
\hcop, \hcccn\ is generally optically thin in galaxies owing to its relatively
low abundance, which makes it an ideal dense gas tracer for calculating the
column density and/or mass of molecular hydrogen content of galaxies.  In the
observations by \citet{Lindberg2011} and \citet{Costagliola2011} a low detection
rate of \hcccn\ was reported and was explained as the intrinsically faint
emission of \hcccn\ , and our stacked result also implies that the \hcccn\ is
quite weak in the non-detected galaxies (2\,$\sigma$ upper limit=0.14 K \kms),
which is also in favor of this explanation.  The non-detection in M 82 is
consistent with the low abundance of \hcccn\ in M 82 suggested by
\cite{Aladro2011}, that \hcccn\ traces a nascent starburst
of galaxy, and it can be easily destroyed by the UV radiation in PDRs, which is
ubiquitous in active galaxies.

In very recent line surveys of a few local active galaxies (AGN and/or
Starbursts, \citealt{Aladro2015}; \citealt{Costagliola2015}), several \hcccn\
transitions in 3mm band (\hcccn\ $J$=10-9, $J$=11-10 and $J$=12-11) were
detected. The ALMA observations by \citet{Costagliola2015} even reported the
\hcccn\ $J$=32-31 rotational transition, and some of the vibrationally excited
\hcccn\ lines.  The latest high resolution line surveys in a few very nearby galaxies
\citep{Meier2005, Meier:2012} and \citet{Meier:2014, Meier:2015} show that, the
derived \hcccn\ abundances (on $\sim$ 100 pc, roughly GMC scales) are about
several 10$^{-10}$ (relative to H$_2$), which is about an order of magnitude
lower than the abundance of HCN and some other molecules.

The results in \citet{Aladro2015} show that, the \hcccn\ fractional abundance is
generally several times lower than that of HCN, \hcop\ and CS. And comparing to
other AGN or starburst galaxies in their sample, \hcccn\ abundance is
significantly higher in the two ULIRGs \object{Arp 220} and \object{Mrk 231},
implying it is suited for studying the activity of ULIRGs. Besides, there was no
obvious evidence of the affection by AGN on the intensity of \hcccn. Four
galaxies in our sample (\object{NGC 1068}, \object{M 82}, \object{Mrk 231} and
\object{Arp 220}) were also studied in \citet{Aladro2015}. We compare our data
with their results, and the \hcccn\ spectra of \object{Mrk 231} and \object{NGC
1068} from \citet{Aladro2015} are shown in Figure \ref{Fig:2_1} and
\ref{Fig:24_23}, to be compared with the non-detection of \hcccn\ 2-1 in
\object{Mrk 231}, and the detection of \hcccn\ 24-23 in \object{NGC 1068},
respectively.  Their results show that, in 3mm band, the intensities between the
three transitions of \hcccn\ (10-9, 11-10 and 12-11) differ not too much, and
the peak temperature ($T_{\rm mb}$) of \hcccn\ are $\sim$ 4 mK for NGC 1068,
$\sim$ 11 mK for M 82, and $\sim$ 1.1 -- 1.7 mK for Mrk 231, and $\sim$ 10 mK
for Arp 220.  In our results, the detection of \hcccn\ 24-23 in NGC 1068 shows a
peak $T_{\rm mb} \sim$ 7 mK, while the non-detection of \hcccn\ 2-1 in Mrk 231
and Arp 220 show that, the RMS we have ($\sim$ 4--6 mK) might not be low enough
to detection the \hcccn\ lines.  Here we conclude that, besides the low
abundance of \hcccn, insufficient integration time and not ideal observing
conditions are the main cause for the low detection rate of \hcccn. 

It would be interesting to compare the intensity ratios between \hcccn\ and
other dense gas tracers, such as HCN and \hcop. We list the ratio between
the flux density of \hcccn\ and HCN 1-0 in Table \ref{Tab:results} and
\ref{Tab:results2}, respectively. Because the data quaity of this work is not
good enough for us to present an accurate estimate on the \hcccn\ flux density,
the ratios are only tentative. We see a large variation in the ratios, which
could be an evidence of the essentially large variation of \hcccn\ luminosities
among galaxies. In the \hcccn\ survey by \citet{Lindberg2011}, ratios like
\hcccn/HCN were used to compare \hcccn\ between galaxies. Based on that ratio,
IC 342 and M 82 were classified as \hcccn-luminous galaxies. In our observation
we detect both \hcccn\ 2-1 and \hcccn\ 24-23 in IC 342, but neither \hcccn\
transition is detected in M 82.  On the other hand, we obtained \hcccn\ 24-23
detections in NGC 1068 which were classified as a \hcccn-poor galaxy in
\citet{Lindberg2011}.  In the sample of some nearby galaxies observed by
\citet{Aladro2015}, the ratio between the peak temperature ($T_{\rm mb}$) of
\hcccn/HCN or \hcccn/\hcop also show large variance. In NGC 253 and M 82,
\hcccn\ 10-9 is only $\sim$ 1/20 as strong as \hcop 1-0, while in Arp 220
\hcccn\ 10-9 is nearly as strong as \hcop 1-0. In our results such line ratios
also show large diversity.  It is not yet clear how to interpret the ratio
between \hcccn\ and other molecular lines, and more data of \hcccn\ in different
transitions would be helpful to disentangle its properties in different types of
galaxies.

Our observations and other works have presented detection of \hcccn\
emission lines from near 18 GHz up to $\sim$ 292 GHz. The newly commissioned
Tianma 65 m telescope in Shanghai, China, is able to observe low
transition \hcccn\ emission, and has great potential for further \hcccn\ 2-1
surveys for large sample of galaxies.

%----------------------------------------------------------------------------

\section{Summary}
\label{sect:summary}
We carry out single-dish observations towards a sample of nearby gas-rich
galaxies with the Effelsberg telescope and the Submillimeter Telescope.
This is the first measurements of \hcccn\ 2-1 in a relatively large sample of
external galaxies. Our results include:
\begin{enumerate}

\item \hcccn($J$=2-1) ($\nu$ = 18.196 GHz) was observed with the 100-m telescope
  in 20 galaxies and only three galaxies are detected ($> 3 \sigma$): \object{IC
  342}, \object{M 66} and \object{NGC 660}.  This is the first measurements of
  \hcccn\ 2-1 reported in external galaxies, and the first \hcccn\ detection in
  M 66. We stack the spectra of those non-detections yet there is still no sign
  of \hcccn\ emission. The $2 \sigma$ upper limit of \hcccn\ intensity from the
  stacked spectrum is about 0.12 K \kms.

\item \hcccn($J$=24-23) ($\nu$ = 218.324 GHz) was observed in nine galaxies with
  the SMT, and it is detected in three galaxies: \object{IC 342}, \object{IC
  694} and \object{NGC 1068}. 

\item IC 342 is the only galaxies detected in both \hcccn\ 2-1 and \hcccn\ 24-23
  transitions in our observations, and the two transitions have similar line
  center and width, suggesting a similar emitting area. The ratio of
  integrated intensity of \hcccn\ 24-23/\hcccn\ 2-1 is about 0.82. Due to the
  contamination of CO 2-1 image signal in the upper sideband, M 82 and Arp 220
  are treated as non-detection of \hcccn\ 24-23.
  
\item The ratios between \hcccn\ and HCN, HCO+ show a large variance among the
  galaxies with \hcccn\ detections, implying different behavior of the molecular
  lines in galaxies. More sample are needed to better understand the
  relationship between \hcccn\ and other molecules.

\end{enumerate}

\begin{acknowledgements} 
 %The authors are very grateful to the anonymous referee
 %for her/his careful reading and constructive comments which improved the
 %paper significantly. 
  We thank the staff of the Effelsberg
  telescope and the SMT for their kind help and support during our observations.
  This project is funded by China Postdoctoral Science Foundation
  (grant 2015M580438), National Natural Science Foundation of China (grant
  11420101002, 11311130491, 11590783 and 11603075), and the CAS Key Research 
Program of Frontier Sciences.
%  and the National Basic Research Program (973 program No.2007CB815405). 
  This research has made use of
  NASA's Astrophysics Data System, and the NASA/IPAC Extragalactic Database
  (NED), which is operated by the Jet Propulsion Laboratory, California
  Institute of Technology, under contract with the National Aeronautics and
  Space Administration.  
\end{acknowledgements}

\appendix                  %%appendicial material is supported

% \section{This shows the use of appendix}
% A postscript file is actually an ASCII text file (you may even edit it).
% However, you need to transfer a PDF file or any compressed or packaged
% file in binary mode when using FTP.
% 
% \section{What is SCI?}
% SCI is the abbreviation of Science Citation Index system powered by
% the Institute for Scientific Information (ISI). For details please
% visit {\it http://apps.isiknowledge.com}.

\bibliographystyle{aa}
\bibliography{jiang_hc3n}

\label{lastpage}

\end{document}